\documentclass[10pt,letterpaper,twocolumn]{article} 

\usepackage{ol2}
\usepackage[draft]{hyperref}
\usepackage{amsmath}

\begin{document}

\twocolumn[ 

\title{Exceptional points and photonic catastrophe}


\author{Stefano Longhi}

\address{Dipartimento di Fisica, Politecnico di Milano and Istituto di Fotonica e Nanotecnologie del Consiglio Nazionale delle Ricerche, Piazza L. da Vinci 32, I-20133 Milano, Italy (stefano.longhi@polimi.it)}

\begin{abstract}
 Exceptional points (EPs) with a global collapse of pairs of eigenfunctions are shown to arise in two locally-coupled and spatially-extended optical structures with balanced gain and loss. Global collapse at the EP deeply changes light propagation, which becomes very sensitive to small changes of initial conditions or system parameters, similarly to what happens in models of classical or quantum catastrophes. The implications of global collapse for light behavior are illustrated by considering discrete beam diffraction and Bloch oscillation catastrophe in coupled waveguide lattices.
\end{abstract}

 ] 

Parity-time ($\mathcal{PT}$) and non-Hermitian photonics are rapidly emerging research areas with a wealth of potential applications \cite{r1,r2,r3}. The ability of creating
and superposing non-Hermitian eigenstates through optical gain
and loss in engineered photonic media has suggested new ways to propagate, control and confine light.
Many of such new routes can be traced to
the existence of  non-Hermitian degeneracies, also known as exceptional points (EPs) \cite{r4,r5,r6}, and spectral singularities \cite{r7,r8,r9,r10}, which are peculiar to 
non-Hermitian classical and open quantum systems \cite{r11}. EPs are singular points in parameter space of a non-Hermitian  Hamiltonian where two
(or more) eigenvalues and their corresponding eigenstates coalesce under a system parameter
variation. EPs have found a wide variety of applications in photonics, including unidirectional transparency \cite{r8,r12,r13,r14}, laser mode control \cite{r15,r16,r17,r18,r19}, 
light structuring \cite{r20}, laser-absorber devices \cite{r9,r21}, optical sensing \cite{r22,r23,r24}, asymmetric mode switching \cite{r25}, and light stopping \cite{r26}. Other interesting behaviors observed near 
EPs are the jamming anomaly \cite{r26bis}, i.e. the drop of the transport through the gain-loss interface as the gain-loss parameter is
increased, and anomalous power law scaling \cite{r26tris}.
The simplest optical system showing an EP is the $\mathcal{PT}$-symmetric dimer \cite{r3,r27}, i.e. a two-level system describing 
mode coupling in two waveguides or resonators, one with gain and the other one with balanced loss.
In spatially-extended systems, EPs appear usually as isolated points in the spectrum of a non-Hermitian Hamiltonian. However, recent works have shown that photonic crystals
 can support rings or contours of EPs in the band spectrum under appropriate symmetries \cite{r29,r30}. Such an eigenfunctions collapse yields significant control over the band structure design and can result in new behaviors, such as the $\mathcal{PT}$ superprism effect and all-angle supercollimation \cite{r30}.\par
In this Letter we suggest a rather general route toward the realization of EPs with a global collapse of eigenfunction pairs, based on generalized  $\mathcal{PT}$-symmetric dimer. We consider two identical spatially-extended optical structures $\mathcal{S}_1$ and $\mathcal{S}_2$, each of one being described by the same Hermitian Hamiltonian $\hat{H}_0$ which depends on space $x$ and, in the most general case, even on time $t$. In the time-independent case, we indicate by $\mathcal{E}_0$ the energy spectrum of each system. The idea of generalized $\mathcal{PT}$-symmetric dimer is realized  by introducing a spatially-homogeneous Hermitian coupling $\sigma$ and balanced spatially-uniform gain and loss $\gamma$ in the two structures [Fig.1(a)]. If the gain/loss strength $\gamma$ is smaller than the coupling $\sigma$, the energy spectrum of the coupled structures remains real and splits into the doublet $\mathcal{E}_{\pm } =\mathcal{E}_0 \pm \sqrt{\sigma^2-\gamma^2}$. As $\gamma$ approaches $\sigma$, the doublet coalesces, corresponding to a global collapse of pairs of eigenfunctions. Note that, while in previous works \cite{r29,r30} collapse of pairs of eigenfunctions is partial,  in our case the collapse can be, at least in principle, complete. A main requirement for the collapse to be complete is that all modes of the structures are coupled with the same strength $\sigma$. Non-Hermitian eigenfunction collapse deeply changes light propagation in the system, for example how light diffracts or localizes in the coupled structures. Like in the theory of classical or quantum catastrophe \cite{r32,r33,r34}, light behavior becomes extremely sensitive to small changes of initial conditions or system parameters: in other words, a  photonic catastrophe is realized near EP operation. The implications of the photonic catastrophe are illustrated by considering discrete beam diffraction and catastrophic Bloch oscillations in waveguide lattices.\par
 \begin{figure}[htb]
\centerline{\includegraphics[width=8.4cm]{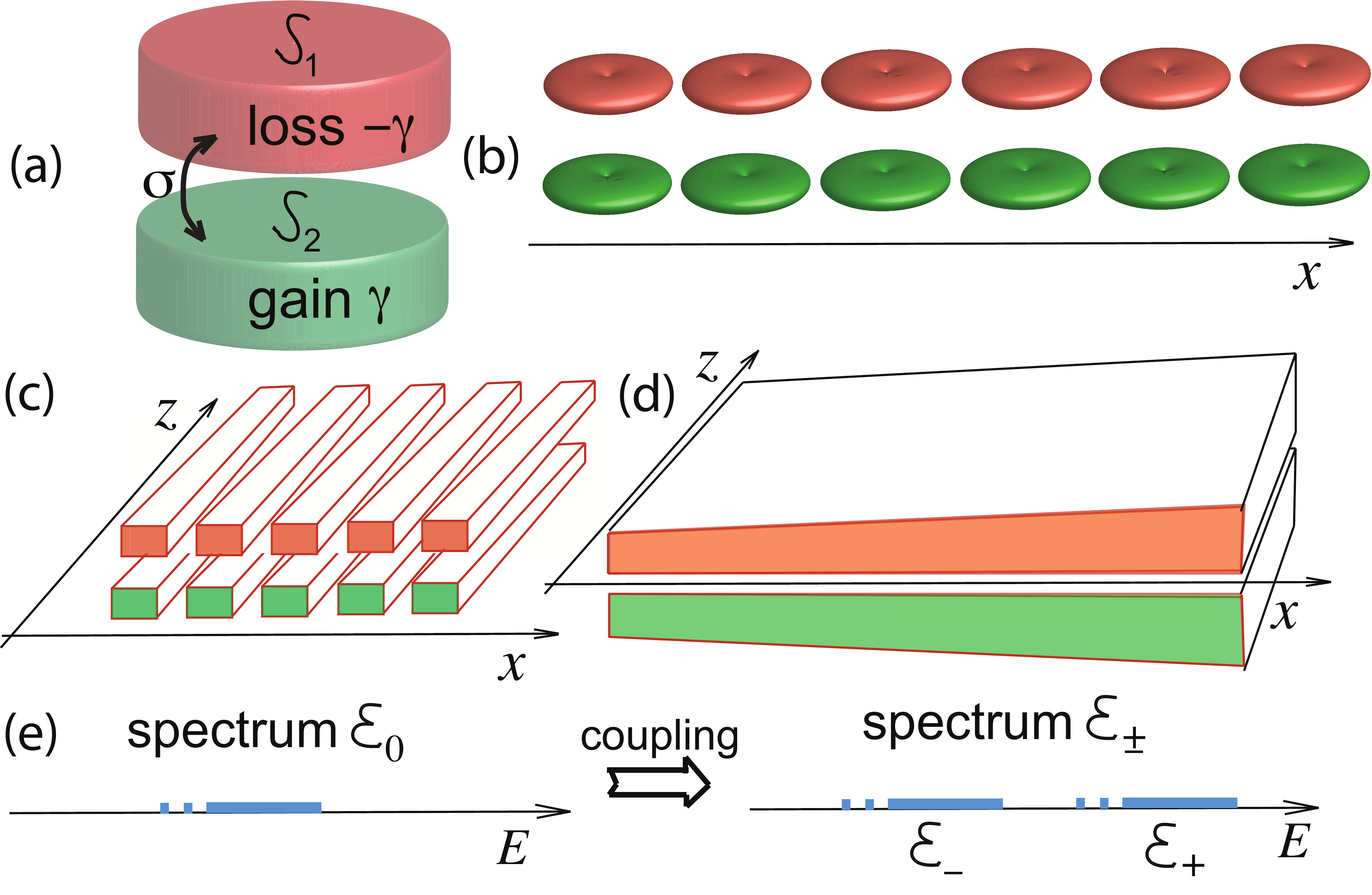}} \caption{ \small
(Color online) (a) Schematic of a generalized $\mathcal{PT}$-symmetric dimer. Two identical optical structures $\mathcal{S}_1$ and $\mathcal{S}_2$ are spatially coupled and provided with homogeneous optical loss ($\mathcal{S}_1$) and balanced gain ($\mathcal{S}_2$). (b-d) Possible realizations of a generalized $\mathcal{PT}$-symmetric dimer: (b) side-coupled CROWs, (c) coupled waveguide lattices, and (d) coupled slab waveguides. (e) The energy spectrum (propagation constant of bound and scattered modes) of each isolated optical structure is $\mathcal{E}_0$. Once spatially coupled and with balanced optical gain and loss, the energy spectrum splits into the doublet $\mathcal{E}_{\pm}$, separated by $\sqrt{\sigma^2-\gamma^2}$. As the loss/gain rate $\gamma$ approaches the coupling $\sigma$, the doublet coalesces and each point of the spectrum becomes a second-order EP.}
\end{figure} 
 Let us consider two identical optical structures $\mathcal{S}_1$ and $\mathcal{S}_2$, such as coupled-resonator optical waveguides (CROWs) or coupled waveguide/slab structures, in which light dynamics, either in time $t$  or in space $z=t$ , is described by the same Hermitian Hamiltonian $\hat{H}_0(x,t)$, where $x$ denotes a set of spatial or site coordinates. For the sake of definiteness, we will refer to the case where $t=z$ describes the spatial propagation coordinate of the optical structure and light waves are monochromatic. Rather generally, we assume that the Hamiltonian  $\hat{H}_0(x,t)$ can vary along the propagation distance $t$. For a $t$-invariant Hamiltonian, the energy spectrum (i.e. propagation constant of modes)  of $\hat{H}_0$ is  indicated by $\mathcal{E}_0$. We then provide some spatially-homogeneous loss $-\gamma$ to the optical structure $\mathcal{S}_1$ and some balanced spatially-homogeneous gain $\gamma$ to the other structure $\mathcal{S}_2$ [Fig.1(a)]. Moreover, we assume that the two systems can be homogeneously and locally coupled, for example via evanescent wave coupling in a side geometry. Examples of optical structures with spatially-homogeneous and local coupling  are shown in Figs.1(b-d), corresponding to side-coupled CROWs [Fig.1(b)], waveguide lattices [Fig.1(c)] and slab waveguides [Fig.1(d)]. In Figs.1(b) and (c) each waveguide/resonator is assumed single-mode, the spatial coordinate $x$ is discrete and spans over the various sites of the chain, whereas in Fig.1(d) $x$ is the transverse coordinate of the slab. Note that translational invariance along $x$ is not required and the optical potential (effective refractive index) can be a rather arbitrary function of $x$. Indicating by $\psi_1(x,t)$ and $\psi_2(x,t)$ the light waves in the two structures, light propagation in the coupled system is described by the spinor wave equations
\begin{eqnarray}
i \frac{\partial \psi_1}{\partial t} & = & \hat{H}_0(x,t) \psi_1-i \gamma \psi_1+\sigma \psi_2 \\
i \frac{\partial \psi_2}{\partial t} & = & \hat{H}_0(x,t) \psi_2+i \gamma \psi_2+\sigma \psi_1 
\end{eqnarray}
where $\sigma$ is the coupling constant. The explicit form of the Hamiltonian $\hat{H}_0$ depends on the specific optical structures. For CROWs or  waveguide lattices [Figs. 1(b,c)], it is a discretized operator while for slab waveguides [Fig.1(d)] is it a Schr\"odinger-like operator, with a potential which is defined by the geometry (local thickness) of the slab \cite{ruff}.
We note that, while we are considering here a $\mathcal{PT}$-symmetric configuration with balanced gain and loss, one could also consider a purely dissipative system in so-called {\it{ passive}} $\mathcal{PT}$ symmetry \cite{r34bis}, which might be more accessible in experiments \cite{r34bis,r34tris,r34quatris}. For a given initial condition $\psi_1(x,0)=f_1(x)$, $\psi_2(x,0)=f_2(x)$, it can be shown that the solution to Eqs.(1) and (2) is given by
\begin{equation}
\left( 
\begin{array}{c}
\psi_1(x,t) \\
\psi_2(x,t)
\end{array}
\right)= \mathcal{U}(t)  \left( 
\begin{array}{c}
\phi_1(x,t) \\
\phi_2(x,t)
\end{array}
\right)
\end{equation}
where $\phi_{1,2}(x,t)$ describe light evolution in the two decoupled structures with $\sigma=\gamma=0$, i.e. $i \partial_t \phi_{1,2}=\hat{H}_0 \phi_{1,2}$ with $\phi_{1,2}(x,0)=f_{1,2}(x)$, and the $2 \times 2$ fundamental matrix $\mathcal{U}(t)$ is the solution of the linear system characteristic of the $\mathcal{PT}$-symmetric dimer
\begin{equation}
i \frac{d \mathcal{U}}{dt}=
\left( 
\begin{array}{cc}
-i \gamma & \sigma \\
\sigma & i \gamma 
\end{array}
\right)
\mathcal{U.}
\end{equation}
with $\mathcal{U}(0)=\mathcal{I}$  (identity matrix). Note that the above solution holds even if the gain/loss $\gamma$ and coupling $\sigma$ vary with $t$. Equations (3) and (4) show that the dynamical behavior of the coupled system is entirely captured by the two-level $\mathcal{PT}$-symmetric dimer model \cite{r3,r27}, regardless of the complexity of the optical potential and spectrum $\mathcal{E}_0$ of $\hat{H}_0$. This means that all features of EPs known for the two-level $\mathcal{PT}$-symmetric dimer, such as unstable linear growth of optical power, topological state-flip and chirality when encircling the EP by static or dynamic parameter variation, etc., should arise for the generalized $\mathcal{PT}$-symmetric dimer model. Let us focus our attention to a stationary Hamiltonian $\hat{H}_0$. Using Eqs.(3) and (4), the energy spectrum and corresponding eigenstates of the coupled optical system can be readily retrieved from the ones of $\hat{H}_0$. If  $\varphi_0(x)$ is an eigenvector of $\hat{H}_0$ with energy $E_0$, $\hat{H}_0 \varphi_0=E_0 \varphi_0$, it readily follows that $(\sigma, i \gamma \pm \sqrt{\sigma^2-\gamma^2})^T \varphi_0(x)$ are eigenstates of the coupled system with energies $E_0 \pm \sqrt{\sigma^2-\gamma^2}$. This means that the energy spectrum of the coupled system is composed by the doublet $\mathcal{E}_0 \pm \sqrt{\sigma^2-\gamma^2}$, which remains real for $\gamma<\sigma$  [Fig.1(e)]. As the gain/loss rate $\gamma$ approaches the coupling $\sigma$, the doublet and corresponding eigenstates coalesce, thus realizing a complete collapse of pairs of eigenfunctions. Such an eigenfunction collapse can deeply change light propagation in the system, for example how light diffracts or localizes in the coupled structures. Like in the theory of classical or quantum catastrophe \cite{r32,r33,r34}, light behavior becomes extremely sensitive to small changes of initial conditions or system parameters near the $\mathcal{PT}$-symmetric breaking (EP)  transition. In particular, below the symmetry breaking transition coupling of the two systems does not   appreciably modify light dynamics as compared to the uncoupled systems: in fact, at the stroboscopic propagation distances  $t=0,t_S,2t_S,3t_S,...$, with $t_S= 2 \pi / \sqrt{\sigma^2-\gamma^2}$, one has  $\psi_{1,2}(x,t)=\phi_{1,2}(x,t)$, as if the two systems were effectively decoupled. However, as the  EP point is approached the stroboscopic distance $t_S$ diverges, light dynamics in the coupled systems is intertwined and   
a qualitative different behavior is found. In fact, for $\gamma=\sigma$, the solution to Eqs.(1) and (2) reads
\begin{eqnarray}
\psi_{1}(x,t) & = & \phi_{1}(x,t)- \sigma t [ \phi_1(x,t)+i \phi_2(x,t)] \\
\psi_{2}(x,t) & = & \phi_{2}(x,t)-i  \sigma t [ \phi_1(x,t)+i \phi_2(x,t)] .
\end{eqnarray}
indicating that the asymptotic light evolution is dominated by the {\em interference} of states $\phi_1(x,t)$ and $\phi_2(x,t)$.\par
 \begin{figure}[htb]
\centerline{\includegraphics[width=8.4cm]{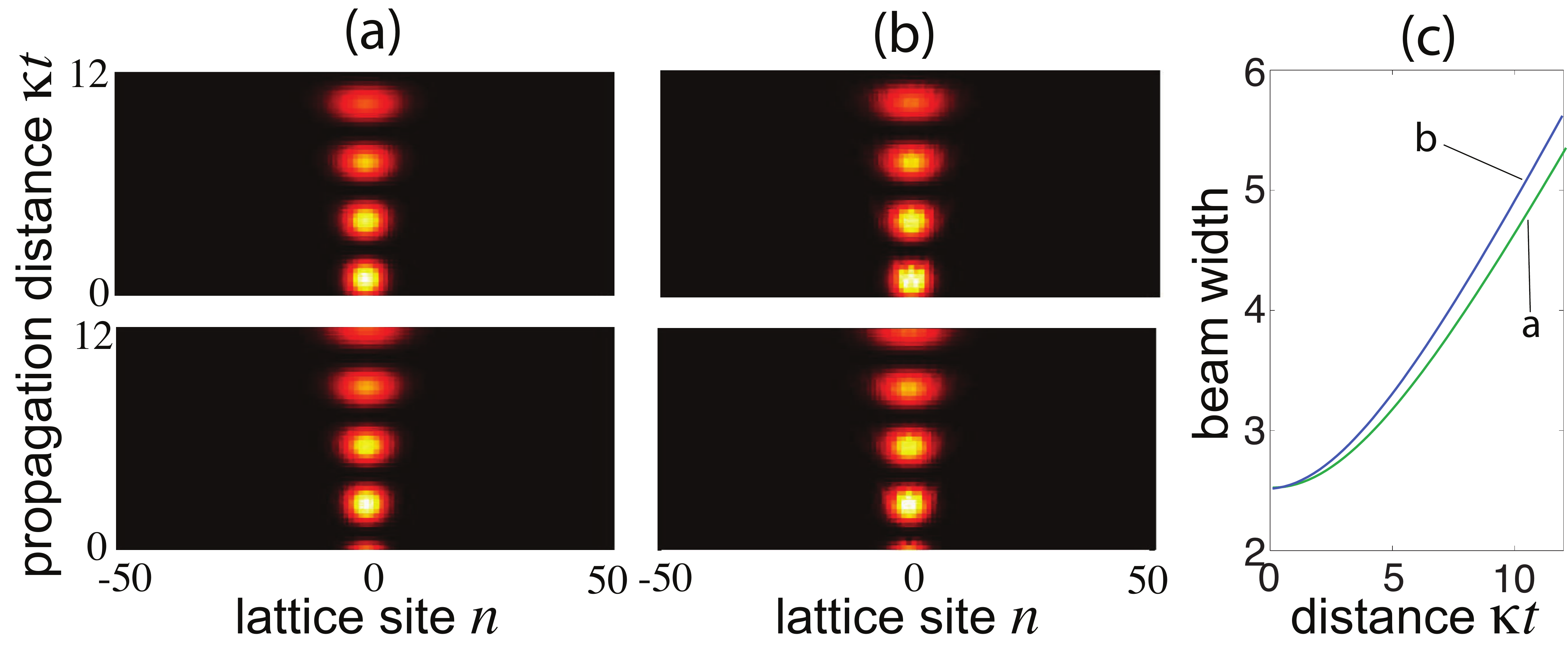}} \caption{ \small
(Color online) (a) Discrete diffraction patterns in the coupled waveguide arrays of Fig.1(c) for $\gamma=0$ and $\sigma=\kappa$. The panels in the figure show the behavior of normalized intensities $|a_n|^2/\sum_n(|a_n|^2+|b_n|^2)$  (upper panel) and $|b_n|^2/\sum_n(|a_n|^2+|b_n|^2)$ (lower panel) on a pseudo color map versus normalized propagation distance $\kappa t$ for Gaussian beam excitation $f(n)=\exp(-n^2/w_0^2)$ with $w_0=5$, as explained in the main text. Rabi flopping between upper and lower arrays, with conserved total optical power, is clearly observed. In (b) the discrete diffraction patterns, corresponding to a slightly different initial condition (a $\pi/10$ phase shift is impressed at site $n=0$ of the lower waveguide lattice), is depicted. Panel (c) shows the evolution of the beam width $w=w(t)$ in the two cases.}
\end{figure} 
 \begin{figure}[htb]
\centerline{\includegraphics[width=8.4cm]{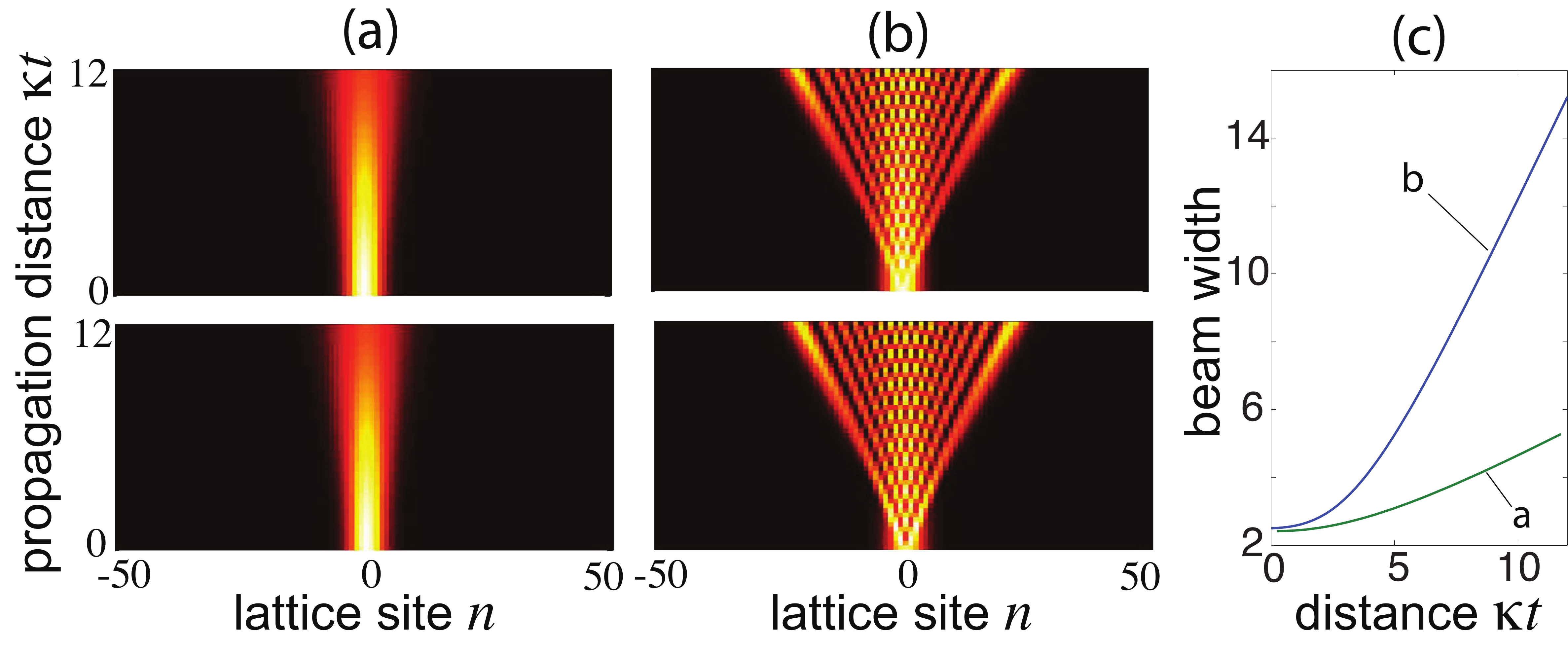}} \caption{ \small
(Color online) Same as Fig.2, but for $\gamma=\sigma=\kappa$ (EP condition). Here total optical power is not conserved.}
\end{figure}
Near the EP, light behavior is expected to undergo a deep qualitative change under small variation of system parameters and/or initial excitation conditions, thus realizing a kind of $^{\prime}$photonic catastrophe$^{\prime}$. Catastrophe theory arises in several areas of classical and quantum physics (see, e.g. \cite{r32,r33}); it studies phenomena characterized by sudden and dramatic shifts in behavior arising from small changes in parameters or initial conditions. Typical examples include change of long-run stable equilibrium states in bifurcation theory \cite{r33} and orthogonal catastrophe in many particle quantum systems \cite{r32}. The interplay between EP and catastrophe theory was earlier suggested in \cite{r34}. Also, a qualitative change of transport behavior near EP, referred to as jamming anomaly, was predicted in \cite{r26tris}. 
Here we illustrate the onset of photonic catastrophe for diffraction and localization of discretized light beams in waveguide lattices \cite{r37,r38,r39}.
Light propagation in the coupled waveguide lattices shown in Fig.1(c) is described by coupled-mode equations \cite{r37,r38,r39}
\begin{eqnarray}
i \frac{d a_n}{dt} & = &  \kappa (a_{n+1}+a_{n-1})+\sigma b_n-i \gamma a_n-Fn a_n  \\
i \frac{d b_n}{dt} & = &  \kappa (b_{n+1}+b_{n-1})+\sigma a_n+i \gamma b_n-Fn b_n
\end{eqnarray}
where $a_n(t) \equiv \psi_1(x=n,t)$ and $b_n(t) \equiv \psi_2(x=n,t)$ are the mode amplitudes in the upper (lossy) and lower (gain) waveguide lattices, $\kappa$ is the coupling constant between adjacent waveguides in each lattice, and $F$ is a gradient force. Coupled $\mathcal{PT}$ dimeric models described by Eqs.(7) and (8) in the limiting case $F=0$ were earlier studied in \cite{uff1,uff2}.
The first example of photonic catastrophe is provided by a giant change of discrete diffraction arising from a small change of initial condition in the homogeneous lattices. For $F=0$, the coupled arrays sustain two shifted bands with the dispersion relations $\mathcal{E}_{\pm}=\mathcal{E}_0(q) \pm \sqrt{\sigma^2-\gamma^2}$, where $\mathcal{E}_0(q)= 2 \kappa \cos(q)$ in the lattice band of the single array and $q$ the Blosh wave number.  Clearly, a global colescence of the two bands arises at $\sigma=\gamma$. Let us assume that the two waveguide lattices are initially excited by two broad optical beams with the same amplitude profile $f(n)$ but with a $\pi/2$ phase shift, i.e.  $a_n(0) \equiv f_1(n)=f(n)$ and $b_n(0) \equiv f_2(n)=i f(n)$, where $x=n$ is the lattice site. This excitation can be realized, for example, by illuminating the array at the entrance plane by a titled Gaussian beam. In the coupled structure and far from the EP (for example in the Hermitian limit $\gamma=0$), a small deviation of beam excitation in the two lattices, measured by the difference $\delta f(n) \equiv f_1(n)+i f_2(n)$, just yields a small change of the discretized light diffraction pattern as compared to the $\delta f(n)=0$ case. This is shown, as example, in Fig.2, where the two waveguide arrays are excited by a broad Gaussian beam at normal incidence $f(n)= \exp(-n^2/w_0^2)$ and the small change $\delta f(n)=[1- \exp(i \theta)] \delta_{n,0}$ is realized by impressing a small phase slip $\theta=\pi/10$ at the $n=0$ lattice site. Clearly, the small change of initial conditions does not appreciably change the discrete diffraction patterns in the two arrays, which show in both cases a rapid Rabi-like flopping superimposed to a slow beam diffraction. In particular, the evolution of beam width $w(t)$, defined as $w^2(t)=\sum_n n^2(|a_n|^2+|b_n|^2)/\sum_n (|a_n|^2+|b_n|^2)$ and shown in Fig.2(c), is slightly modified by the perturbation of initial excitation condition. However, at the EP ($\gamma=\sigma$) the same small change of initial condition yields a huge change in the discrete diffraction patterns, as shown in Fig.3. The reason thereof is that, while for exact excitation $f_2(x)=if_1(x)$ the unstable growing mode in Eqs.(5) and (6) vanishes because of exact destructive interference, a small deviation of the excitation condition yields imperfect destructive interference and the unstable mode will dominate the dynamics. For the small change $\delta f(n)=[1- \exp(i \theta)] \delta_{n,0}$ of initial condition, the discrete diffraction pattern of dominant unstable mode is the impulse response of the array at site $n=0$, resulting in an enhanced  discrete diffraction, i.e. in a $^{\prime}$superdiffraction$^{\prime}$; see Fig.3(c) right panel. The example shows that an extremely small change of initial condition (for example  of the phase in one lattice site) can be exploited to engineer discrete diffraction. Other kinds of initial condition perturbations, for example a small change of the phases of two adjacent sites, could produce other behaviors, such as giant beam refraction. Also, considering more than two coupled arrays, one could in principle realize global eigenfunction collapse in the spectral domain corresponding to a higher-order EP, e.g. third-order EP for three arrays. In this case the algebraic unstable growth at the EP \cite{r26bis} could make light diffraction or refraction even more sensitive to a small change of initial conditions. \\
As a second example, let us consider Bloch oscillations (BOs) in the coupled waveguide lattices of Fig.1(c) induced by an index gradient  $F$, which can be realized by circularly-bending the optical axis \cite{r37,r38,r39}. In this case, the Hamiltonian $\hat{H}_0$ of the single array has a purely point spectrum, namely a Wannier-Stark (WS) ladder spectrum with energy levels equally-spaced by $F$. This yields periodic self-reconstruction of the beam after a propagation distance $t_B= 2 \pi /F$, i.e. so-called optical BOs. For the coupled array with $\gamma=0$, a doublet of WS ladders is obtained, which are spectrally shifted relative to each other by the coupling constant $\sigma$. As a result, like in other two-band waveguide lattices the BO dynamics is rather generally quasi-periodic \cite{r40}, but if the ratio $\sigma /F$ is rational perfect reconstruction of the beam is still observed. In particular, if $\sigma$ is an integer multiple than $F$, the two WS ladders coalesce. This Hermitian degeneracy does not obviously spoil the periodic BO motion, which occurs at the period $t_B=2 \pi /F$. This is shown, as example, in Fig.4(a), where BO dynamics in the coupled arrays for initial single-site excitation is depicted for $\sigma / F=1$. For a non-vanishing gain/loss $\gamma$ below the $\mathcal{PT}$ symmetry breaking point, the spacing between the two WS ladders is reduced to $\sqrt{\sigma^2-\gamma^2}$, and a similar scenario (periodic or quasi-periodic dynamics) is found like in the Hermitian case.  However, as the EP $\gamma \rightarrow \sigma^-$ is reached, the two WS ladders coalesce in both energy and eigenstates, which implies the disappearance of any periodic/ quasi-periodic dynamics for generic initial excitation of the arrays. Such a $^{\prime}$BO catastrophe$^{\prime}$, induced by non-Hermitian collapse of WS ladders, is illustrated in Figs.4(b) and (c). 
 \begin{figure}[htb]
\centerline{\includegraphics[width=8.4cm]{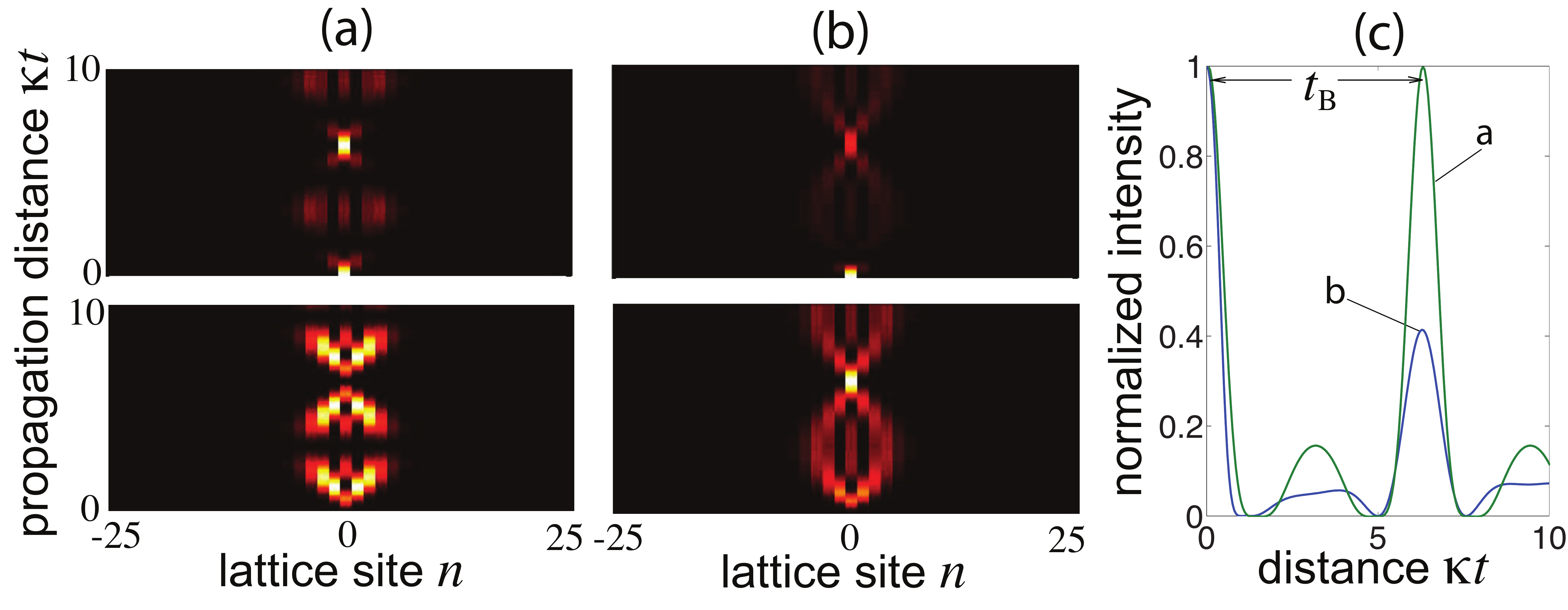}} \caption{ \small
(Color online) BO dynamics in coupled waveguide lattices for $\kappa=\sigma=F=1$ and for (a) $\gamma=0$, and (b) $\gamma=\sigma$. Panels (a) and (b) depict the evolution, along the normalized propagation distance $\kappa t$, of the normalized intensity distributions $|a_n|^2/ \sum_n (|a_n|^2+|b_n|^2)$ (upper panels) and $|b_n|^2/ \sum_n (|a_n|^2+|b_n|^2)$ (lower panels) in the two waveguide arrays for single-site excitation of the upper array $a_{n}(0)=\delta_{n,0}$, $b_n(0)=0$.  Panel (c) shows the detailed behavior of the normalized intensity distribution at initially-excited lattice site in the two cases. In (a) there is a Hermitian collapse of the two WS ladders, which leads to periodic BO dynamics with spatial period $t_B= 2 \pi/F$. In (b) there is a non-Hermitian collapse of the WS ladders, which spoils BOs.}
\end{figure}
  
In conclusion,  EPs with a global collapse of pairs of eigenfunctions arise when two spatially-extended optical structures with balanced gain and loss are locally coupled. Like in the theory of catastrophes, close to EP light propagation becomes very sensitive to small changes of initial conditions or system parameters. This behavior might be of interest for  applications in optical control and optical sensing. The implications of  non-Hermitian eigenfunction collapse have been illustrated by considering discrete beam diffraction and Bloch oscillation catastrophe in coupled waveguide lattices. Here we have considered static and linear EP operation, however it would be interesting to extend the study to dynamical and/or nonlinear regimes, for example in EP encircling problems, where further intriguing  behaviors could be found. Also, collapse corresponding to EPs of higher-order, providing a higher sensitivity of system behavior to parameter changes, could be investigated.

\end{document}